# Strain Induced Modulation of Local Transport of 2D Materials at the Nanoscale

R. Maiti[1,⊥], M.S.A.R. Saadi[2,⊥], R. Amin[1], V. O. Ozcelik[3], B. Uluutku[2], C. Patil[1], C. Suer[1], S. Solares[2], V. J. Sorger[1],*

[1]Department of Electrical and Computer Engineering, George Washington University, Washington, DC 20052, USA
[2]Department of Mechanical and Aerospace Engineering, George Washington University, Washington, DC 20052, USA
[3]Materials Science and Engineering Program, University of California at San Diego, San Diego, California, USA
*sorger@gwu.edu

**Abstract:**
**Strain engineering offers unique control to manipulate the electronic band structure of two-dimensional materials (2DMs) resulting in an effective and continuous tuning of the physical properties. Ad-hoc straining 2D materials has demonstrated novel devices including efficient photodetectors at telecommunication frequencies, enhanced-mobility transistors, and on-chip single photon source, for example. However, in order to gain insights into the underlying mechanism required to enhance the performance of the next-generation devices with strain(op)tronics, it is imperative to understand the nano- and microscopic properties as a function of a strong non-homogeneous strain. Here, we study the strain-induced variation of local conductivity of a few-layer transition-metal-dichalcogenide using a conductive atomic force microscopy. We report a novel strain characterization technique by capturing the electrical conductivity variations induced by local strain originating from surface topography at the nanoscale, which allows overcoming limitations of existing optical spectroscopy techniques. We show that the conductivity variations parallel the strain deviations across the geometry predicted by molecular dynamics simulation. These results substantiate a variation of the effective mass and surface charge density by .026 $m_e$/% and .03e/% of uniaxial strain, respectively, derived using band structure calculation based on first principle density functional theory. Furthermore, we show and quantify how a gradual reduction of the conduction band minima as a function of tensile strain explains the observed reduced effective Schottky barrier height. Such spatially-textured electronic behavior via surface topography induced strain variations in atomistic-layered materials at the nanoscale opens up new opportunities to control fundamental material properties and offers a myriad of design and functional device possibilities for electronics, nanophotonics, flextronics, or smart cloths.**

**Introduction:**

Strain engineering of 2D materials (2DMs) has attracted attention, both scientifically and technologically, given its ability to sustain a large amount of strain as compared to bulk semiconductors [1-6]. In contrast, bulk three-dimensional (3D) materials are often associated with low failure strain (about 0.1%) due to the presence of lattice defects and dislocations thus limiting their applications in 'strain(op)tronics'. On the other hand, strain in 2D materials has demonstrated generation and modulation of intriguing electrical, optical, and mechanical properties, such as improving electrical mobility or dramatically altering the band gap, for instance, thus showing promise for next-generation functional devices [6,7].

Strain in 2DMs can be achieved by several techniques including (a) bending and stretching of flexible substrates, (b) relaxation of pre-stretched substrates, (c) substrate surface topography modification, (d) tip indentation of suspended 2DMs, and (e) piezoelectric substrate actuation. This is in contrast to their bulk 3D counterparts where strain is typically applied by forcing the epitaxial growth of heterojunctions with a lattice parameter mismatch [6-13]. Among these different techniques, understanding and quantifying the amount of strain generation in 2DMs induced by substrate topography is critical for subsequent device engineering-[10-12]. Therefore, strain in thin 2DMs can be manipulated by shaping the attributes of these underlying structure (size, shape, height, and density). One of the examples is an 'artificial atom', where, transferring a CVD-grown $MoS_2$ monolayer on lithographically patterned nanocones, gives rise to spatially tunable bandgap capable of broadband light absorption [14]. Moreover, placing and thereby augmenting 2DMs onto photonic integrated circuit (PIC) structures, such as waveguide or micro-ring resonator, offers i) seamless heterogenous integration without any required lattice matching issues known form compound semiconductors, and ii) strong modulation of physical properties by external perturbation such as, electric field, optical excitation, or strain. Indeed, 2DMs-PIC integration allows for a plurality of functionality including an alteration in a change in the carrier density, polaritonic states, atomistic Stark-effect, helical topological states [15-17]. Thus, such induced material property changes the complex refractive indices of the material, and tunes the optical response enabling active functional devices. For instance, using stain as an external 'knob' one can engineer the bandgap in order to enhance the performance of optoelectronic devices, exemplary in photodetectors and emitters [6,7,18,19].

Thus far, a plethora of fascinating physical phenomena of 2DMs induced by strain have been presented, such as the modulation of the bandgap [6, 19, 20], direct-indirect bandgap transition [21], semiconductor-metallic phase change [22], softening or hardening of phononic modes [23], electron–phonon coupling alteration [24], to share exemplary work. However, most of these properties have been studied using the optical spectroscopic techniques such as micro-Raman and photoluminescence spectroscopy [12, 19-21, 25]. These commonly employed techniques to estimate or derive strain cannot represent the entire picture for 2D material based nanoscale devices since they provide average spatial information restricted by the diffraction-limited laser spot size, with typical beam diameters being on the order of micrometers. Therefore, at this stage, when nanoscopic devices are experiencing an unprecedented upsurge in both academic research and practical applications, understanding the electronic and optical properties of the materials as a function of non-homogeneous strain directly at the nanoscale becomes a prerequisite for designing strain-engineered devices and thus provides the original impetus for the present study.

Here, we investigate the local conductivity of a few-layer $MoTe_2$ when integrated onto a step like ridge geometry as a function of strain using a Conductive Atomic Force Microscope (CAFM). The variation of conductivity matches the strain build up across the ridge as calculated by Molecular Dynamics (MD) simulation and can be explained by the variation of mobility and surface charge as studied by density functional theory (DFT) based band structure calculations. Our local electronic transport measurements on top of spatially tunable strained device allow us to not only to understand the strain-induced bandgap modulation, but also provides insights into a local strain map, thus shedding light onto device junction physics such as providing a quantitative analysis of the effective Schottky barrier height, as demonstrated here. Together, these findings provide key experimental evidence for highly tunable electronic properties at high resolution, which constitutes a new paradigm of nanoscale strain engineering in 2DMs with particular relevance for developing performance strainoptronic devices.

**Results & Discussions:**

In order to probe the charge transport phenomena as a function of strain, we perform local electrical measurements using a conductive AFM tip scanning across a step-like ridge structure across a vertical M-S-M device which is realized by the precise placement capability of the exfoliated $MoTe_2$ flakes by utilizing our in-house-developed 2D material printer [26] technique atop a thin (10 nm) Au layer that acts as a bottom electrode on top of a silicon-on-insulator photonic waveguide (**Fig. 1a-b**). Thus, henceforth, we refer to the local strain-inducing surface topology of 'waveguide', yet the concept is universal just depending on the underlying topology. This allows obtaining locally-resolved strain-induced bandgap modulation due to its influence on the electrical conductivity. After the topography is imaged in tapping mode (3D topography, **Fig. 1c**) to avoid surface damage and tip wear, leveraging the precise tip motion and location ability of AFM, *I-V* curves were collected at different locations across the waveguide (*I-V* locations marked on the topography in **Fig. 1c**). A voltage is applied to the tip and current passing through the 2D nanocrystal and a bottom gold-coated substrate collects the current using a DLPCA-2000 amplifier and recorded using the ARC2 AFM controller. The same tip and contact forces (~50 nN) are used for all the measurements to ensure similar experimental conditions. The corresponding local *I-V* results (**Fig. 1d**) suggest a graded modulation of conductivity in the material across the waveguide (**Fig. 1e**) and show a maximized conductivity near the edges of the waveguide, a reduced value atop, and minima away from the waveguide.

This observed modulation in electrical conductivity can be understood in terms of bandgap modulation induced due to a locally-varying strain across the waveguide [6]. At room temperature, in the subthreshold regime, the conductivity of a semiconductor can be expressed as [9], $\sigma_{strained} = \sigma_{unstrained} * \exp[-\frac{\varepsilon}{2kT*Eg}*\frac{\partial Eg}{\partial \varepsilon}]$, where, $\sigma_{strained}$ and, $\sigma_{unstrained}$ stand for the conductivity of MoTe$_2$ flakes for the strained and unstrained region, respectively. $k$ is the Boltzmann constant, $T$ the temperature, $\varepsilon$ the strain, and $\partial Eg/\partial \varepsilon$ the variation of the bandgap as a function of strain. The enhanced conductivity is a result of two effects, namely, the reduction of the bandgap of the semiconductor (piezoresistive effect) and the modulation of the Schottky barrier (piezotronic effect). In the subthreshold regime, where transport is dominated by thermally excited carriers, the curves are symmetric and linear suggesting that the transport behavior is dominated by the piezoresistive effect rather than piezotronic effect [27, 28]. In order to estimate the graded strain generated in the material due to the underlying structure, a bandgap of few layer 2H MoTe$_2$ was calculated using DFT and plotted as a function of the applied strain. A binomial fit to the data is used to extract the strain dispersive bandgap term (**Fig. S2**, supplementary information). Taking the region considerably away from the waveguide as the unstrained region, we can estimate the amount of local strain across the waveguide to be about 3-4% (**Fig. 1e**). We find that the so experimentally measured strain follows the trend obtained by molecular dynamics (MD) simulations, thus denoting a consistent data set (**Fig. 1f**).

Furthermore, in order to understand the mechanism of strain generation and its spatial variation due to the underlying substrate geometry, we recreate the structure with the same structural proportionality (but reduced dimensions for computational tractability) to perform MD simulation, thus extracting the strain variation across the geometry based on the comparison of the atomic positions in the MoTe$_2$ tri-layer before and after energy minimization of the structure (**Fig. S3**, supplementary online information). An average strain mapping (average of three layers) displays an overall tensile trend with a maximum in the suspended region near the edges of the waveguide (along the tensile direction within the 2D layers) (**Fig. 1f**). Attractive van der Waals forces between the suspended region and the substrate causes the material to stretch slightly in the entire suspended region to allow a larger portion of its distal ends to remain closer to the substrate, thus generating strain.

This approach would require rearrangement and reconfiguration of Mo-Mo and Mo-Te bond lengths to accommodate the generated strain. This miniaturized calculation is not expected to quantitatively reproduce the strain of the experimentation precisely as (1) the region of favorable van der Waals interactions in the simulation is small (these interactions decay rapidly with distance anyway), and (2) the separation between the 2D layers and the substrate in the miniaturized structure increases rapidly away from the distal ends; that is, under the Euler-Bernoulli beam formalism, the molybdenum atomic layer in MoTe$_2$ acts as the natural axis of the structure and since the strain calculation is performed based on the displacement of the atoms in the central molybdenum strip (which are considered to be the most representative for strain calculations), most of the bending effects in a single layer is negated. However, unlike the single layer structure, bending effects are much more substantial in multi-layer structures and may be up to 4 times that of the strain in the principle direction. At the very edges of the sheet(s) some strain concentration can be observed as well, which is attributed to the edge relaxation mechanism of mechanically exfoliated 2D materials [29].

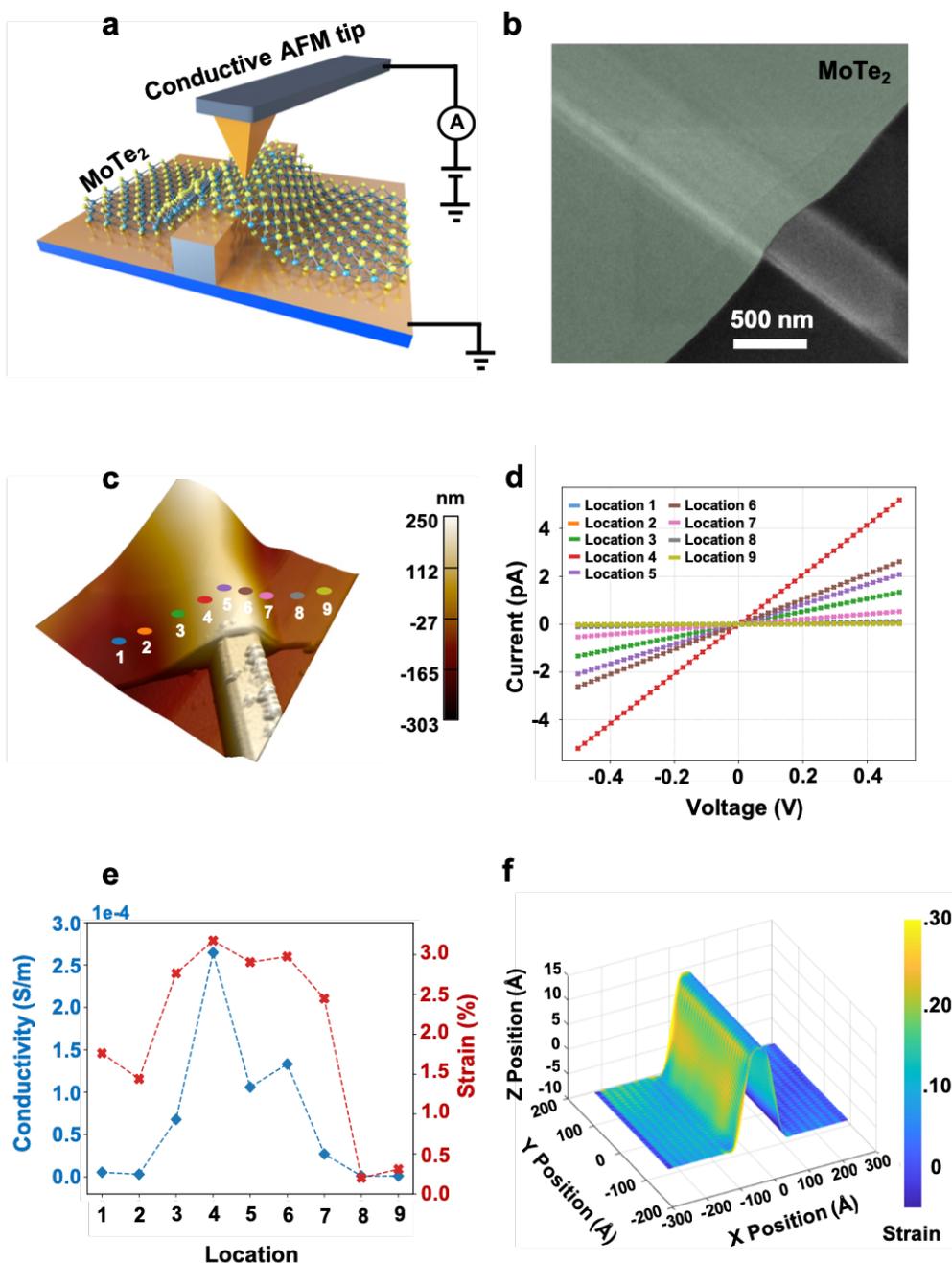

**Figure 1: Spatially tunable Local conductivity** (a) schematic representation of the measurement setup where local conductivity is probed by CAFM on MoTe$_2$ wrapped around a ridge-like structure (e.g. waveguide). (b) False color SEM micrograph of strained MoTe$_2$ layer (thickness-30 *nm*)wrapped around a silicon-on-insulator step-like ridge. Ridge parameters; *h* = 220 nm; *w*=500 nm). (c) AFM topography of the device where *I-V* curves were collected at 9 different locations, as shown, around the waveguide. (d) *I-Vs* at a small bias window (subthreshold regime), where transport is mostly dominated by piezoresistive effect than piezotronic effect. (e) Conductivity and locally varying strain calculated from the graded conductivity plotted against different locations. (f) An average strain map for a 3-layer MoTe$_2$ using molecular dynamics (MD) simulation showing a maximum strain at the edges of the waveguide and on the suspended section of the flake.

To gain further insights into the conductivity change induced by the spatially varying non-homogeneous strain, we explore its impact on the electronic band structure by performing density functional theory (DFT) calculations. We identify the optimized geometries of various MoTe$_2$ monolayer structures under external strain and calculated their electronic band diagrams for each case. The positions of the atoms in the unit cell and unit cell dimensions were reoptimized for each strain value to find the optimized atomic configuration for each case by minimizing the internal stress and force on each atom. This method was previously shown to be successful at accurately capturing the electronic and structural properties of similar two-dimensional materials under external strain [30,31].

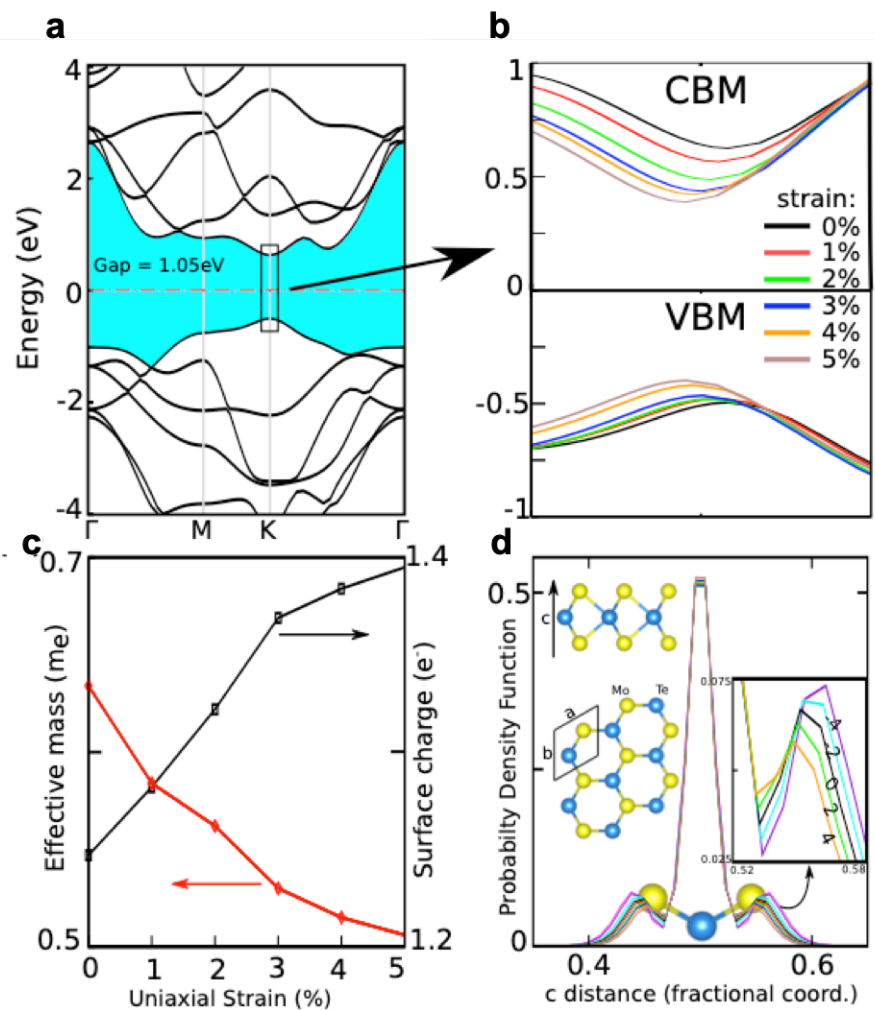

**Figure 2. Evolution of electronic band structure of MoTe$_2$ under strain,** (a) Band structure of pristine MoTe$_2$ calculated with Generalized Gradient Approximation (GGA) method with vdW corrections. The Fermi level which is set to E=0 is shown with red dash-dotted line. (b) Variation of the CBM and VBM states as uniaxial strain is varied between 0 – 5%. (c) Variation of effective mass of charge carriers and surface charge density with uniaxial strain. (d) Probability density function at CBM along the out-of-plane c-direction of MoTe$_2$, where Mo atoms are at c=0.5. The inset on the right shows a zoomed-in section of the plot near the surface Te atoms for different compressive and tensile strains values of -4, -2, 0, 2, and 4% where the negative sign indicates uniaxial compression of the unit cell.

The electronic structure and energy bands of the pristine monolayer MoTe$_2$ structure show a fundamental bandgap of 1.05 eV (**Fig. 2a**). Both the valance band maximum (VBM) and the conduction band minimum (CBM) reside on the K high-symmetry point of the Brillouin zone and denotes a direct bandgap semiconductor. In our DFT model, we apply an external uniaxial strain (upto 5%) to monitor the evolution of the CBM and VBM states as well as the distribution of free electrons along the material as a function of strain. We find that as the strain value increases from 0 to 5%, the CBM and VBM states move closer to the Fermi level (which is set to 0 eV) and the band gap gradually decreases to 0.77 eV for 5% uniaxial strain (**Fig. 2b**) [6]. This corresponds to a 27% drop in the band gap as compared to the pristine structure which also affects the surface carrier density of the MoTe$_2$ layer. To compute this effect, we next calculate the variation of the effective mass of electrons as a function of strain.

At the band edge in the Brillouin Zone (the K high symmetry point), the electronic energy changes linearly with respect to $\mathbf{q} = K - \mathbf{k}$, where $\mathbf{q}$ is a value corresponding to a point $\mathbf{k}$ in the vicinity of the K point. This leads to $E = \hbar \mathbf{v}_F q + O[(\mathbf{q}/\mathbf{k}^2)]$, where $\mathbf{v}_F$ is the Fermi velocity. Therefore, by calculating the derivate of the electronic energies of the bands near the K-point with respect to $\mathbf{q}$, the Fermi velocity of the charge carriers can be computed for each strain value. Similarly, using the curvature value of CBM around the K-point, the effective mass can estimated by $\hbar/m^* = 1/(d^2E/dk^2)$ (**Fig. 2c**). Accordingly, the effective mass decreases from 0.64 to 0.51m$_e$ (25% change), where m$_e$ is the mass of an electron in vacuum. Since lower effective mass increases the mobility of electrons, they are more free to move toward the surface under stain, which also increases the surface carrier density. In addition, the surface carrier density increases from 1.24*e* to 1.39*e* (11% change). These values indicate an increase in the mobility of surface charge carriers as external uniaxial stress increase to 5% (**Fig. 2c**).

To evaluate the origin of the increased mobility of charge carriers in strained MoTe$_2$, we can find the real space wavefunction along the out-of-plane direction of the single layer structure under different strain values. The wavefunction is a proxy for the spatial distribution of electrons which helps to understand the origin of charge movement inside the material. Here, we are interested in the out-of-plane direction to monitor the movement of electrons toward the surface of MoTe$_2$. Using the Generalized Gradient Approximation (GGA) scheme, we obtain the plane wave coefficients of each lattice point in the reciprocal space of the unit cell, from which the wavefunctions were constructed. Using the pseudo-wave function, $\psi_{n,k}(r)$, of a specific band (n) at the K-point, we can extract the probability function corresponding to a specified region in the real space as $\int_{z_1-z_2} dz \int_\Omega \psi_{n,k}(r)^2 dxdy$, where $\Omega$ is the two dimensional cross section of the unit cell, and $z_1$-$z_2$ are the boundaries in the out-of-plane direction where the wavefunction is evaluated. Here, the wavefunctions are computed at the K-point where the CBM resides (**Fig. 2d**). We find that as strain value increases, the wavefunction spreads along the c-axis towards the surface atoms and the barrier for the electrons to move from Mo to surface Te atoms decrease, thus supporting the argument for a higher surface charge density under strain.

Next, we investigate the effect of locally varying strain on the tip-MoTe$_2$ (metal-semiconductor) junction. Since, charge transport in 2D materials strongly depends on the Schottky barrier height (SBH), therefore understanding its role and modulation will facilitate the optimization of device performance [32]. For example, in case of a photodetector, low contact resistance in 2D/Metal contact is critical to obtain high a gain-bandwidth product [33]. Previously, we have demonstrated that the 2DM's band structure can be engineered by using strain as an external perturbation [6, 10, 20]. However, the effect of geometry induced local strain to the modulation of SBH in the MoTe$_2$/metal (conductive AFM tip) contact has not been systematically and comprehensively studied so far. Here, we explore the effect of uniaxial tensile-strain on the SBH of a few-layered MoTe$_2$ flake through vertical Au/MoTe$_2$/Au junctions using an Gold-coated AFM tip (the substrate is also gold coated). However, the major issue for 2DM-based devices is the existence of a large barrier height at the 2D/metal junction resulting in high contact resistance values, which usually limits device performance (e.g. RC-delay speed, energy dissipation). Such high resistance originates primarily from the absence of dangling bonds in pristine 2DMs which drastically influence charge injection and limits the device for high frequency operation [6, 34]. On the other hand, as the SBH approaches zero (ohmic contact), carriers of the 2DM can propagate to metal layers almost freely. Therefore, controlling the SBH is of importance for any device applications concerning 2D materials.

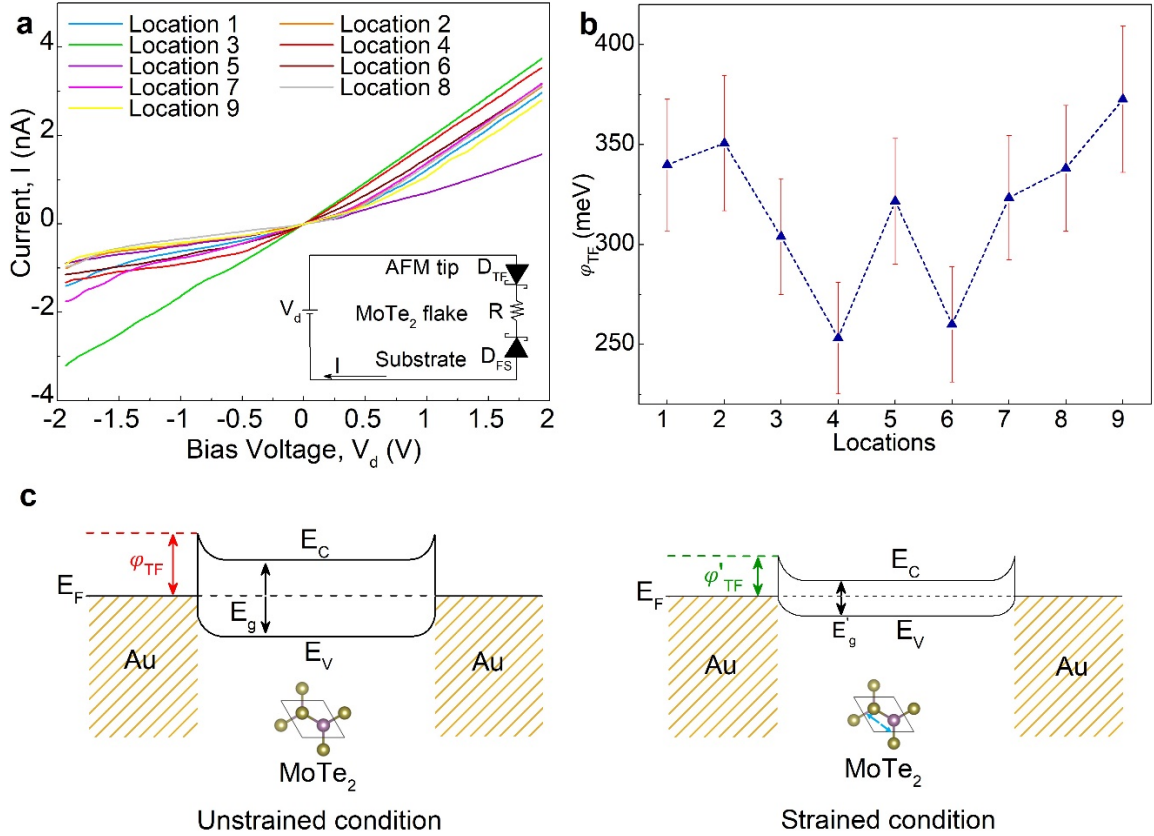

**Figure 3. Schottky barrier alteration with uniaxial strain at metal contacts.** (a) Conductive atomic force microscopy (CAFM) achieved *I-V* characteristics for 9 different locations across the underlying waveguide structure corresponding to differently strained regions. An equivalent circuit model is shown as an inset for the measurement arrangement showcasing the back to back Schottky diode model for the Au AFM tip, $D_{TF}$ and Au substrate junctions, $D_{FS}$ and the MoTe$_2$ flake as a resistor, $R$. (b) Effective AFM tip-MoTe$_2$ flake junction barrier height, $\varphi_{TF}$ (meV) for the aforementioned 9 different locations across the structure exhibiting strain dependent barrier lowering trends. (c) Schematic energy band representation of the Au-MoTe$_2$-Au setup showing unstrained (outside the structure) and strained (structure localized) conditions exhibiting the strain induced lowering of the barrier energy from $\varphi_{TF}$ to $\varphi'_{TF}$.

When a metal is deposited on a semiconductor, it generally creates a Schottky contact inducing a potential barrier ($\phi_B$) depending upon the work function of the metal ($\phi_m$) and the electron/hole affinity ($\chi_s$) of the semiconductor. Therefore, according to Schottky-Mott rule, SBH = q($\phi_m$-$\chi_s$), which suggests choosing a metal with a low work function with the Fermi level aligned close to the conduction band of the 2D material, will facilitate electron injection, whereas, a high work function with the Fermi level aligned close to the valence band of the 2D material allows for hole injection. For example, Wei et al. showed that monolayer MoS$_2$ interfacing with Ir, Pd, and Ru leads to SBH of 0.66 eV, 0.79 eV, and 0.72 eV respectively [35].

Our local I-V characteristics measured using CAFM probe show a rectifying behavior where two metal–semiconductor junctions are formed at the tip/MoTe$_2$ and MoTe$_2$/substrate interfaces (**Fig. 3a**). The variation of SBH indicates a lower barrier height as a function of strain, i.e. extrema at locations 4 and 6, corresponding to the edge of the waveguide edge location. This variation of SBH corroborates well with conductivity map (**Fig. 1e**) suggesting that the enhanced conductivity as a function of tensile strain mainly originates from two factors: i) strain induced bandgap lowering, and ii) alterations of SBH at the contacts. In order to determine the contribution from SBH variation due to the uniaxial tensile strain, here we analyze our *I-V* data for different locations across the geometry by using a two Schottky diode connected back to back equivalent circuit (**Fig. 3a, inset**). The current flowing in the device can be modelled according to the thermionic emission theory as [36-37],

$$I_i(V_i) = I_{0,i} \exp\left(\frac{qV_i}{n_i k_B T}\right)\left[1 - \exp\left(-\frac{qV_i}{k_B T}\right)\right] \quad (1)$$

$$I_{0,i} = A_c A^* T^2 \exp\left(\frac{q\varphi_i}{k_B T}\right) \quad (2)$$

where, $I_i$ is the current through the junction upon application of bias voltage, $V_i$; $I_{0,i}$ is the saturation current of the junction, $q$ is the electronic charge, $n_i$ is the ideality factor of the diode, $k_B$ is the Boltzmann constant, $T$ is the temperature, $A_c$ is the metal–semiconductor contact area, $A^*$ is the effective Richardson constant and $\varphi_i$ is the energy of the Schottky barrier. The tip-flake (TF) and flake-substrate (FS) interfaces can be modelled as Schottky diodes connected in a back-to-back formation with the semiconductor side of the junction facing opposite directions and the series resistance, $R$, between the two diodes accounts for the finite conductivity of the MoTe$_2$ flake (**Fig. 3a, inset**). The measured *I–V*s in and their least-square fits to the theoretical model (see supplementary information) are fitted with multiple parameters including the series resistance, $R$, the ideality factors, $n_i$, and the saturation currents, $I_{0,i}$ following methods adopted in recent literature and our results closely follow similar trends, e.g. lowering of the barrier height with applied uniaxial strain [38-40]. Transport characteristics through both the diodes and the resistor are influenced by the applied strain. The flake-substrate barrier ideality and saturation current are found to be relatively unaltered without any uniaxial strain around 1.01 and 51 nA, respectively. The tip-flake barrier ideality factor is found as rather consistent at approximately 1.03 with the associated saturation current varying closely following the conductivity traits across the waveguide structure (**Fig. 1c**). The variation of the saturation current (see supplementary information) is associated with the alteration of the Schottky barrier formed at the tip-flake interface probably due to a strain induced change of the MoTe$_2$ electron affinity, as has been reported for similar 2D materials [9, 19, 22]. The saturation current has an exponential dependence on the barrier [41].

The change in both the sub-threshold regime device conductance and the saturation currents attained from full bias range I-Vs suggest that not only the device conductance but also the potential barriers formed at the electrical contact regions are modified by strain. The effective Richardson coefficient is taken as $A^* = 4\pi m^* q k_B^2/h^3 = 1.20173 \times 10^6 \, Am^{-2}K^{-2} \times (m^*/m_e)$ where $m^*/m_e$ are taken from the DFT calculations corresponding to the strain variations on the MoTe$_2$ flake draped waveguide structure (**Fig. 2c**). The extracted effective SBH corresponding to the CAFM tip and the strained flake, $\varphi_{TF}$ exhibit a reduction in energy with the increase in uniaxial strain augmented by corresponding locations in the suspended sections of the MoTe$_2$ flake from the corners of the waveguide ridge (**Fig. 3b**).

While these results are attained with a combination of experimental data and a perturbative heuristic approach involving numerical solutions for the oppositely polarized Schottky diodes in the equivalent circuit, they only represent general trends of the barrier with strained positions, since there are many mechanisms which can cause non-ideal Schottky barrier diode behavior, such as thin interfacial insulating layer between the metal tip and the flake, parasitic oxide charge around the anode, rough or damaged flake surface, tunneling current, etc. [42]. Note, the schematic band diagram (**Fig. 3c**) represents strain induced lowering of the barrier height arising from our underlying waveguide structure following similar traits portrayed in recent literature [39,40]. Incidentally, in this context, we have recently shown the associated shrinking of the bandgap and the increase of work function of the TMD with applied strain (**Fig. 3c and Fig. S5**) [6].

In conclusion, we studied the local electronic behavior of a few layer transition-metal-dichalcogenide two-dimensional material as a function of spatially varying uniaxial tensile strain. Furthermore, we demonstrate a novel characterization technique based on Conductive Atomic Force Microscopy to show key material property modification as a function of local structural topology inducing altered material strain. Our findings not only demonstrate a novel strain characterization method featuring nanoscale resolution, but more generally show and quantify, exemplary for this material system, how local topology induces tensile strain impacting transport and optical material parameters such as effective mass, mobility, bandgap, and Schottky barrier height. This demonstrated novel material property control of the Schottky barrier height as a function of strain, enables added functionality in 2D materials-based Van der Waals heterostructures for applications in next-generation electronic and photonic devices [43-46].

## Methods

### MD simulation

The molecular dynamics (MD) simulations were performed using LAMMPS software (version, 19 September 2019) [R1]. Intra-crystal interactions of $MoTe_2$ were calculated using Stillinger-Weber (SW) potential. [R2-R8] Non-bonded interatomic interactions between the substrate-$MoTe_2$ and between $MoTe_2$ layers are calculated using Lennard-Jones (LJ) potential. LJ parameters used in the simulation is calculated using Lorentz-Berthelot mixing rules from the values presented in table 1[R9]. The silicon substrate is assumed to be rigid.

**Table 1:** LJ parameters for atoms to calculate the interaction between the substrate and layers of the flake.

| Material | $\epsilon$ (meV) | $\sigma$ (Å) |
|---|---|---|
| Te-Te[R10] | 14.3 | 4.428 |
| Si-Si[R11] | 25.4 | 3.385 |
| Mo-Mo[R12] | 0.585 | 4.2 |

The simulation is conducted using a two-step energy minimization process, as in our previous work [R13]. To summarize the process; first an initial structure is created by using additional supporting layers attached to silicon substrate around the representative wave guide. Second, the $MoTe_2$ is fixed, clamped, from both ends and another energy minimization is performed after removing the supporting structures. A representative image before minimization and the dimensions of the structure can be found in the figure S3. During energy minimization, $10^{-20}$ error tolerance is used. After the second minimization, strain calculations for individual $MoTe_2$ layers are done according to the molybdenum atoms of the layers.

### DFT Simulation

DFT results were obtained by performing first principles calculations within generalized gradient approximation (GGA) including van der Waals corrections [R14]. We used projector-augmented wave potentials [R15] and approximated exchange correlation potential with Perdew-Burke-Ernzerhof functional [R16]. We sampled the Brillouin zone (BZ) in the Monkhorst-Pack scheme where a k-point sampling of 21x21x1 was found to be suitable for the BZ corresponding to the primitive unit cell. The energy convergence value between two consecutive steps was chosen as $10^{-6}$ eV. Numerical calculations were carried out using the VASP software [R17].